%% file: main.tex
\documentclass[aps,prl,twocolumn,10pt,amsmath,amssymb,bibnotes,superscriptaddress,longbibliography]{revtex4-1}

\usepackage{graphicx, braket}
\usepackage{siunitx}
\usepackage[colorlinks,urlcolor=blue,citecolor=blue,linkcolor=blue]{hyperref}
\newcommand{\is}{\ensuremath{I_{CS}}}
\newcommand{\vp}{\ensuremath{V_D}}
\newcommand{\vc}{\ensuremath{V_T}}
\newcommand{\di}{\ensuremath{\Delta I_{CS}}}
\newcommand{\g}{\ensuremath{\Gamma}}
\newcommand{\eo}{\ensuremath{\epsilon_0}}
\newcommand{\gt}{\ensuremath{\Gamma/k_B T}}
\newcommand{\dndt}{\ensuremath{dN/dT}}
\newcommand{\zeroone}{\ensuremath{0 \to 1}}
\usepackage{xcolor}

\usepackage[normalem]{ulem}

\begin{document}

\title{Entropy measurement of a strongly coupled quantum dot}

\author{Timothy Child}
\email{timjchild@phas.ubc.ca}
	\affiliation{Stewart Blusson Quantum Matter Institute, University of British Columbia, Vancouver, British Columbia, V6T1Z4, Canada}
	\affiliation{Department of Physics and Astronomy, University of British Columbia, Vancouver, British Columbia, V6T1Z1, Canada}
\author{Owen Sheekey}
	\affiliation{Stewart Blusson Quantum Matter Institute, University of British Columbia, Vancouver, British Columbia, V6T1Z4, Canada}
	\affiliation{Department of Physics and Astronomy, University of British Columbia, Vancouver, British Columbia, V6T1Z1, Canada}
\author{Silvia L\"{u}scher}
	\affiliation{Stewart Blusson Quantum Matter Institute, University of British Columbia, Vancouver, British Columbia, V6T1Z4, Canada}
	\affiliation{Department of Physics and Astronomy, University of British Columbia, Vancouver, British Columbia, V6T1Z1, Canada}
\author{Saeed Fallahi}
	\affiliation{Department of Physics and Astronomy, Purdue University, West Lafayette, Indiana, USA}
	\affiliation{Birck Nanotechnology Center, Purdue University, West Lafayette, Indiana, USA}
\author{Geoffrey C. Gardner}
	\affiliation{Birck Nanotechnology Center, Purdue University, West Lafayette, Indiana, USA}
    \affiliation{School of Materials Engineering, Purdue University, West Lafayette, Indiana, USA}

\author{Michael Manfra}
	\affiliation{Department of Physics and Astronomy, Purdue University, West Lafayette, Indiana, USA}
	\affiliation{Birck Nanotechnology Center, Purdue University, West Lafayette, Indiana, USA}
	\affiliation{Elmore Family School of Electrical and Computer Engineering,  Purdue University, West Lafayette, Indiana, USA}
    	\affiliation{School of Materials Engineering, Purdue University, West Lafayette, Indiana, USA}
\author{Andrew Mitchell}
    \affiliation{School of Physics, University College Dublin, Belfield, Dublin 4, Ireland}
    \affiliation{Centre for Quantum Engineering, Science, and Technology, University College Dublin, Dublin 4, Ireland}
\author{Eran Sela}
    \affiliation{Raymond and Beverly Sackler School of Physics and Astronomy, Tel-Aviv University, IL-69978 Tel Aviv, Israel}
\author{Yaakov Kleeorin}
	\affiliation{Center  for  the  Physics  of  Evolving  Systems,  University  of  Chicago,  Chicago,  IL,  60637,  USA}
\author{Yigal Meir}
	\affiliation{Department of Physics, Ben-Gurion University of the Negev, Beer Sheva 84105, Israel}
	\affiliation{The Ilse Katz Institute for Nanoscale Science and Technology, Ben-Gurion University of the Negev, Beer Sheva 84105, Israel}
\author{Joshua Folk}
\email{jfolk@physics.ubc.ca}
	\affiliation{Stewart Blusson Quantum Matter Institute, University of British Columbia, Vancouver, British Columbia, V6T1Z4, Canada}
	\affiliation{Department of Physics and Astronomy, University of British Columbia, Vancouver, British Columbia, V6T1Z1, Canada}
\date{\today}

\begin{abstract}

The spin 1/2 entropy of electrons trapped in a quantum dot has previously been measured with great accuracy, but the protocol used for that measurement is valid only within a restrictive set of conditions.  Here, we demonstrate a novel entropy measurement protocol that is universal for arbitrary mesoscopic circuits and apply this new approach to measure the entropy of a quantum dot hybridized with a reservoir, where Kondo correlations typically dominate spin physics.    The experimental results match closely to numerical renormalization group (NRG) calculations for small and intermediate coupling.  For the largest couplings investigated in this work, NRG predicts a suppression of spin entropy  at the charge transition due to the formation of a Kondo singlet, but that suppression is not observed in the experiment.

\end{abstract}

\maketitle
Entropy is a powerful tool for identifying exotic quantum states that may be difficult to distinguish by more standard metrics, like conductance. For example, bulk entropic signatures in twisted bilayer graphene indicate that carriers in some phases with metallic conductivity retain their local moments, as would normally be associated with a Mott insulator \cite{Saito.2021, Rozen.2021, Lian.2021}. Entropy has also been proposed as a tell-tale characteristic of isolated non-abelian quasiparticles, whether Majorana modes in a superconductor \cite{Smirnov.2015r9,  Sela.2019} or excitations of a fractional quantum Hall state \cite{cooper2009observable,PhysRevB.79.115317,Ben-Shach.2013}, distinguishing them from abelian analogs.

Quantifying the entropy of single quasiparticles is challenging due to the small signal size, of order $k_B$, but first steps in this direction have been made in recent years~\cite{Hartman.2018,Kleeorin.2019}. Ref.~\onlinecite{Hartman.2018} employed Maxwell relations
to measure the $k_B \ln{2}$ spin entropy of a single electron confined to a quantum dot (QD) in GaAs via the temperature-induced shift of a Coulomb blockade charge transition.  That approach relied on the assumption of weak coupling between the QD and the reservoirs, in order to fit based on the specific charging lineshape known for that regime. In that weak-coupling regime, spin states are pristine enough to serve as spin qubits \cite{Elzerman.2004, Petta.2005, Hanson.2007, Barthel.2009, Bluhm.2010, Nowack.2011, Shulman.2012} but the underlying physics is very simple.

The weak-coupling approach of Ref.~\onlinecite{Hartman.2018} is not applicable to a broad class of mesoscopic devices\cite{Pyurbeeva.2021x0i}, which limits its value in probing the complex Hamiltonians that may be implemented in such systems. For example, a single-impurity Kondo effect is realized when the localized spin is strongly coupled to a reservoir \cite{Goldhaber-Gordon.1998,Pustilnik.2004}. Recently, more complicated structures including multiple dots have been engineered to host multi-channel Kondo states \cite{Potok.2007, Keller.2015}, a three-particle simulation of the Hubbard model \cite{Dehollain.2020}, and more.  Entropy measurements made on any of these systems would offer a significant advance in their understanding, but are not yet possible.

Here, we develop a universal protocol for mesoscopic entropy measurement that forgoes the simplifying assumptions of Ref.~\onlinecite{Hartman.2018}, then apply it to investigate the entropy of the first electron as it enters a quantum dot when strongly hybridized with a reservoir, the regime where Kondo correlations are expected. The protocol is based on a Maxwell relation expressed in integral form,
\begin{equation}\label{eq:eran}
    \Delta S_{\mu_{1}\to \mu_{2}} = \int_{\mu_{1}}^{\mu_{2}}\frac{dN(\mu)}{dT}d\mu,
\end{equation}
\noindent that provides access to the entropy change, $\Delta S$, with chemical potential $\mu$, based on measurements of the change in quantum dot occupation, $N$, with temperature, $T$ \cite{Sela.2019}.  
When the coupling, $\Gamma$, between dot and reservoir is weak ($\Gamma\ll k_B T$), the characteristic temperature of Kondo correlations is much less than the experimental temperature, $T_K\ll T$, and the measurement matches well to single-particle approximations.  When $\Gamma\gtrsim k_B T$, the onset of entropy as the electron enters the dot is strongly modified.  The measurement of this modified entropy signature is the primary result of this work.

\begin{figure}
    \includegraphics[width=1.0\columnwidth]{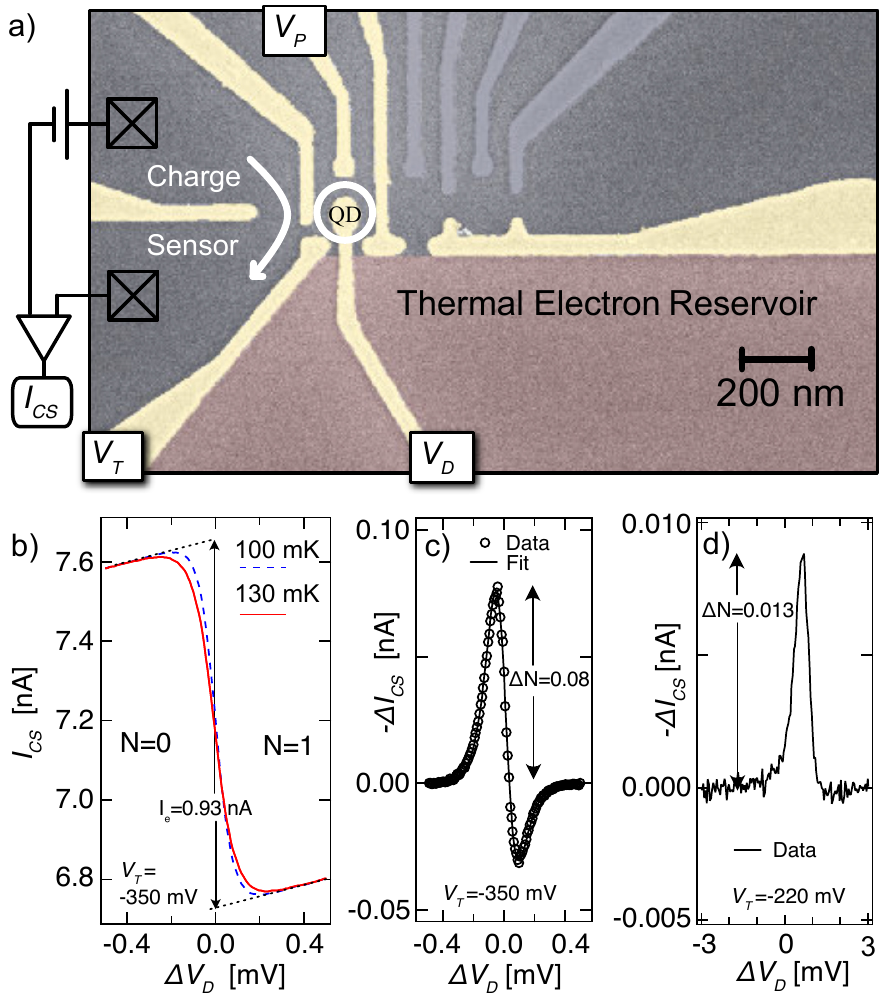}
    \caption{\label{fig:fig1} 
    a) Scanning electron micrograph (colour added) of the device. Electrostatic gates (gold) define the circuit in a 2D electron gas. Grey gates were unused and remained grounded. Squares represent ohmic contacts to the 2DEG. The thermal electron reservoir (red) was alternated between base and elevated temperatures. b) Current through the charge sensor for the \zeroone~ charge transition in a weakly coupled regime, separated into the unheated (100mK) and heated (130mK) parts of the interlaced measurement \cite{methodspaper}, showing $N=0\rightarrow 1$ step height $I_e$. c,d) Change in $I_{CS}$ from 100 to 130 mK, for weak (c) and strong (d) coupling between QD and reservoir.  c) includes fit to weakly-coupled theory, yielding $\Delta S_{\rm{fit}}=(1.02\pm 0.01)k_B \ln 2$. 
    }
\end{figure}

Measurements were performed on a circuit (Fig.~\ref{fig:fig1}a) defined by voltages applied to electrostatic gates over a GaAs 2D electron gas \cite{supplement}, similar in functionality to that in Ref.~\onlinecite{Hartman.2018}, with improvements made in the design and entropy measurement protocol \cite{methodspaper}.
The circuit includes the QD, a charge sensing quantum point contact, and an electron reservoir that can be rapidly Joule-heated above the chip temperature $T$ to an elevated $T+\Delta T$. Both $T$ and $T+\Delta T$ were calibrated by fitting to thermally broadened charge transitions; except where noted,  $T=100$ mK with $\Delta T \sim 30$ mK.
Coupling between the QD and the thermal reservoir is via a single tunnel barrier, with transmission controlled by \vc.

Coarse adjustments in the QD energy level were made using gate voltage $V_P$, with fine adjustments controlled by \vp.  In practice, $V_P$ and \vp~tune the chemical potential as it enters into Eq.~\ref{eq:eran} by adjusting the difference in energy between $\mu$ in the reservoir and the energy needed to add an electron to the dot\cite{Pyurbeeva.2021x0i}.  Throughout this paper we report \vp~with respect to to the midpoint of the charge transition, $\Delta \vp\equiv \vp -\vp(N=1/2)$.

$N$ was monitored via the current, \is, through the nearby charge sensor \cite{Elzerman.2004}, which was biased with a DC voltage between 50 and 300 \textmu V. 
Fig.~\ref{fig:fig1}b illustrates weakly coupled $N=0\rightarrow 1$ transitions at two different temperatures, with $I_e$ defined as the net drop in $\is$  across the transition due to the addition of $1e$ charge to the dot.   Measurements at the two temperatures were interlaced by alternated Joule heating of the reservoir at 25Hz to reduce the impact of charge instability, then averaged over several sweeps across the charge transition, see Ref.~\onlinecite{supplement}.

Figure~\ref{fig:fig1}c shows the difference in the detector current  in the weakly coupled regime  between  100 and 130 mK (shown in Fig.~\ref{fig:fig1}b),  $\Delta\is(V_D)\equiv\is(T+\Delta T,V_D)-\is(T,V_D)$.  Note that $-\Delta\is$ is plotted instead of $\Delta\is$ in order to connect visually with $\Delta N$, which increases when $\is$ decreases.
As in Ref.~\onlinecite{Hartman.2018}, the lineshape of $\Delta\is(V_D)$ in Fig.~\ref{fig:fig1}c may be fit to a non-interacting theory for thermally-broadened charge transitions to extract the change in entropy across the transition, $\Delta S_{\rm{fit}}$, not requiring calibration factors or other parameters  (see Ref.~\onlinecite{Hartman.2018} for details).  For the data in Fig.~\ref{fig:fig1}c, this yields $\Delta S_{\rm{fit}} = (1.02\pm 0.01)k_B \ln{2}$, where the uncertainty reflects standard error among 5 consecutive measurements at slightly different \vc.

The limitation of this approach is illustrated by the very different lineshape in Fig.~\ref{fig:fig1}d, reflecting the $0\to 1$ transition when the QD is strongly coupled to the reservoir and Kondo correlations begin to play a role. Fitting this data to thermally-broadened theory would yield a meaningless $\Delta S_{\rm{fit}}>10 k_B \ln{2}$ for the spin-1/2 electron, motivating the integral approach from Eq.~\ref{eq:eran}.

Before moving to the quantitative evaluation of entropy via Eq.~\ref{eq:eran}, we note that the different lineshapes of $\Delta\is(V_D)$ in Figs.~\ref{fig:fig1}c and d---the peak-dip structure in Fig.~\ref{fig:fig1}c contrasting with the simple peak in Fig.~\ref{fig:fig1}d---can be understood as representing two temperature regimes for the Anderson impurity model.  Fig.~\ref{fig:fig1}c represents the high temperature limit, where $dN/dT$ is approximately a measure of the energy derivative of the density of states in the QD, and thus exhibits positive and negative lobes.
At sufficiently low temperatures, the exact solution \cite{tsvelick1983exact}, and the resulting Fermi liquid theory \cite{mora2015fermi} predict a positive $dN/dT$ for all values of the chemical potential, from the empty level to the Kondo regime through the mixed-valence regime, with a peak expected at a chemical potential corresponding to $T_K(\mu)\sim T$, where the entropy is expected to crossover from $S=0$ to $S=k_B \log2$. Fig.~1d, corresponding to data taken when $T\ll\Gamma$, demonstrate such all-positive $dN/dT$.

For a quantitative extraction of entropy beyond the weakly-coupled regime of Fig.~\ref{fig:fig1}c, we follow the integral approach in Eq.~\ref{eq:eran} that makes no assumptions on the nature of the quantum state.  Evaluating Eq.~\ref{eq:eran} also provides a measurement of $\Delta S(\mu)$ that is continuous across the charge transition, rather than just comparing $N=0$ to $N=1$ values.  To evaluate Eq.~\ref{eq:eran} from experimental data, $dN(\mu)/dT$ is approximated by the ratio $\Delta N(V_D)/\Delta T=-\Delta \is(V_D)/(I_e\Delta T)$.  $\Delta T$ is expressed in units of gate voltage using the corresponding lever arm\cite{supplement} so that the integral may be evaluated over $V_D$, giving $\Delta S(V_D)$.

\begin{figure}
         \includegraphics[width=1.0\columnwidth]{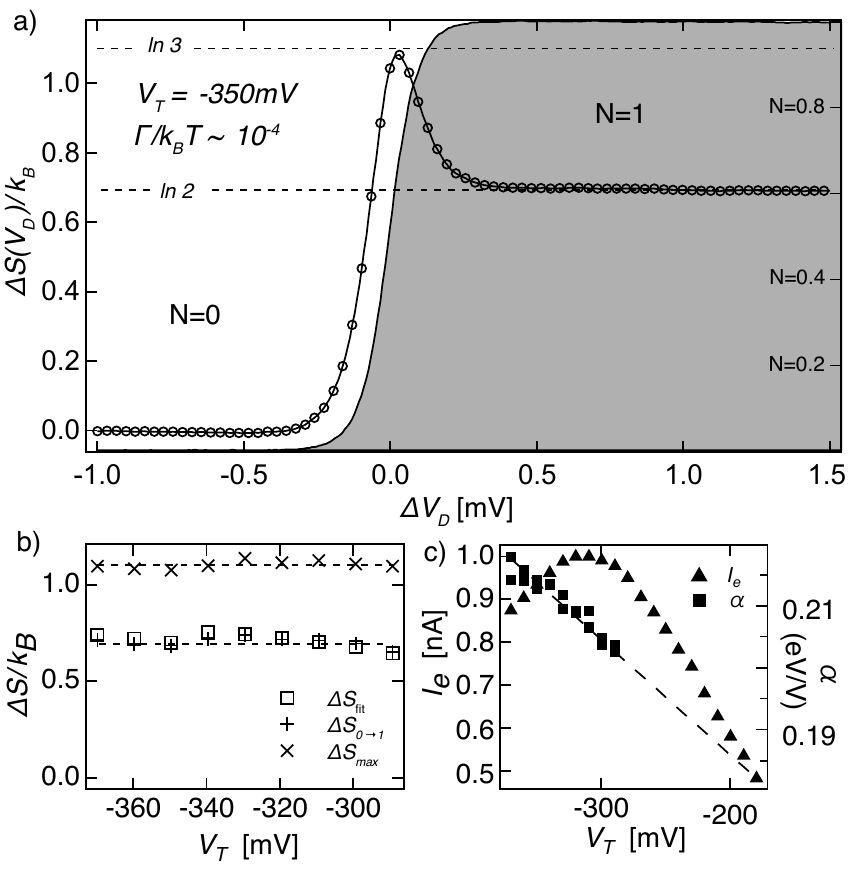}
         \caption{\label{fig:fig2}
         a) Change of $S$ in the QD across the $N=0\rightarrow 1$ transition, obtained by integrating $\Delta\is(V_D)$ (Fig.~\ref{fig:fig1}c) following Eq.~\ref{eq:eran}. Dot occupation across the transition is shown in grey.  Data obtained in the weakly coupled limit, $V_T=-350$ mV corresponding to $\g/k_BT\sim 1\times 10^{-4}$. $\Delta S_{0\rightarrow 1}=(0.99\pm 0.02) k_B \ln 2$ is the net change $\Delta S$ across the complete transition. b) Comparison of $\Delta S_{\rm{fit}}$, $\Delta S_{0\rightarrow 1}$, and $\Delta S_{max}$ (see text) for \vc~covering approximately $10^{-5}<\g/k_B T<10^{-1}$. c) Variation of charge step $I_e$ and lever arm $\alpha$, measured independently over the full range of \vc~explored in this experiment. Dashed line: extrapolation of $\alpha$ into the strongly-coupled regime where it cannot be measured directly. 
         }
 \end{figure}

We begin by confirming the integral approach in the weakly-coupled regime, where the physics is simple.  Fig. \ref{fig:fig2}a shows the entropy change across the $N=0\rightarrow 1$ charge transition for $\Gamma\ll k_B T$, calculated from the data in Fig.~\ref{fig:fig1}b using Eq.~\ref{eq:eran}. The resulting $\Delta S(\mu)$ indicates that the change in dot entropy is non-monotonic as the first electron is added, reaching a $k_B\ln{3}$ peak before settling to $k_B\ln{2}$.  The $k_B\ln{3}$ peak just above $\Delta\vp = 0$ reflects a combination of charge and spin degeneracy on the dot in the middle of the charge transition, with three possible microstates $\{\ket{N=0},\ket{N=1,\uparrow},\ket{N=1,\downarrow}\}$ all equally probable. Charge degeneracy is gone after the transition, but spin degeneracy remains, leaving two possible microstates $\{\ket{N=1,\uparrow},\ket{N=1,\downarrow}\}$. The net change in entropy from beginning to end, $\Delta S_{0\to 1}=(0.99\pm 0.02)k_B \ln{2}$, is nearly identical to the $\Delta S_{\rm{fit}} = (1.02\pm 0.01)k_B \ln{2}$ from Fig.~\ref{fig:fig1}c, despite very different sources of error for the two approaches.  

\begin{figure}
         \includegraphics[width=1.0\columnwidth]{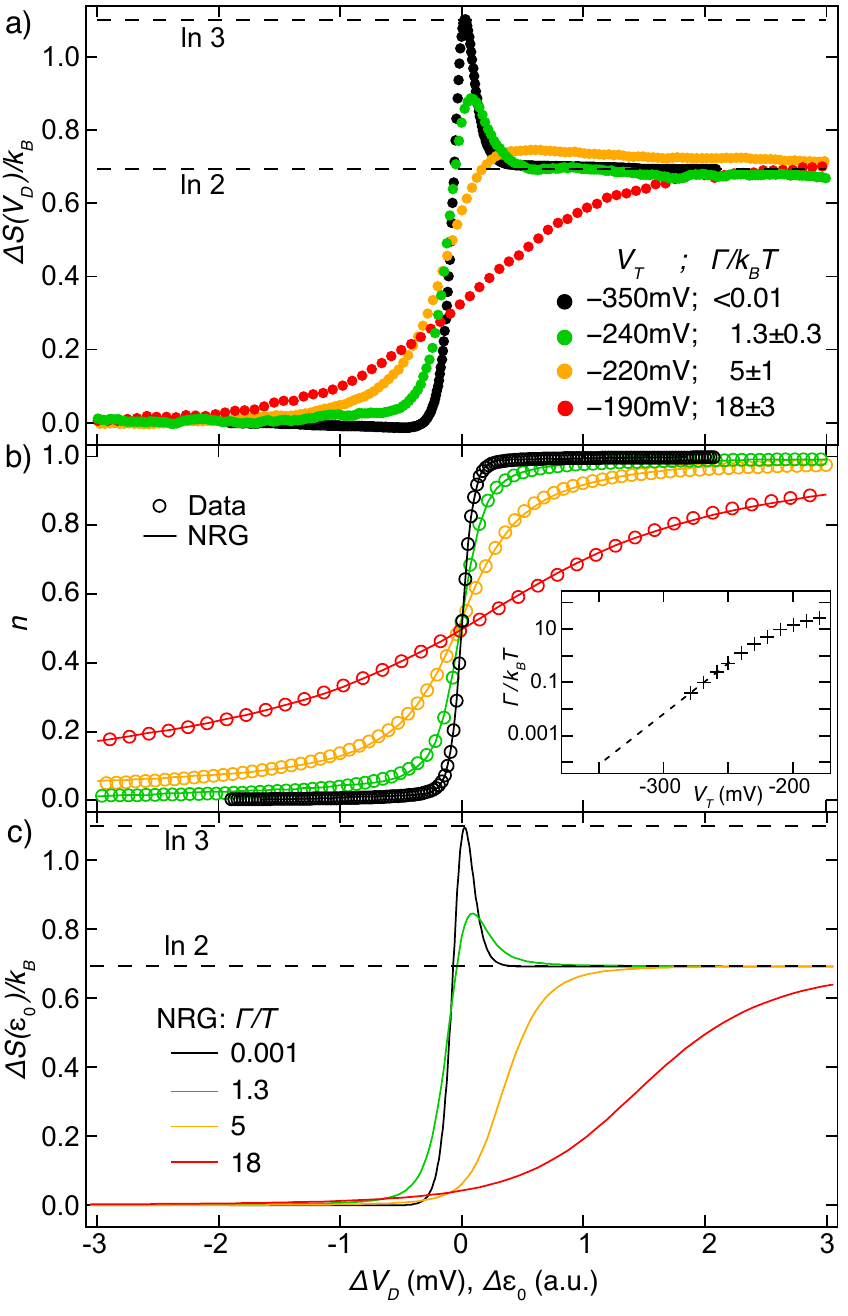}
         \caption{\label{fig:fig3}
         Evolution of $S(\mu)$ from the weak (black) to strong (red) coupling regimes, comparing data (panel a) to NRG calculations (panel c). Measurements of occupation across the charge transition are fit to NRG (panel b), leaving no free fit parameters for the $S(\mu)$ calculation. Panel b inset: Coupling strength of the QD to the reservoir, \gt, extracted from  fits, across the full range of \vc. Values $\gt\ll 1$ cannot be measured directly and are extrapolated (dashed line). 
         }
 \end{figure}

Figure \ref{fig:fig2}b compares the fit and integral approaches for weakly-coupled charge transitions covering four orders of magnitude in $\Gamma$, with the coupling tuned by \vc~(see Fig.~\ref{fig:fig3}b inset for calibration of $\Gamma$).
The consistency between $\Delta S_{0\to 1}$ and $\Delta S_{\rm{fit}}$  over the full range of weakly-coupled \vc, in addition to the fact that $\Delta S_{max}$ remains $k_B\ln$ 3 throughout this regime, confirms the accuracy of the integral approach. Small deviations from $\Delta S_{0\to 1}=\Delta S_{\rm{fit}}=k_B \ln{2}$, such as that seen around \vc = -330 mV, are repeatable but sensitive to fine-tuning of all the dot gates; we believe they are due to extrinsic degrees of freedom capacitively coupled to the dot occupation, such as charge instability in shallow dopant levels in the GaAs heterostructure.

After confirming the accuracy of Eq.~\ref{eq:eran} in the weakly coupled regime, we turn to the regime $\Gamma \gtrsim k_B T$ ($\vc > -280$ mV), where the influence of hybridization and Kondo correlations is expected to emerge. Fig.~\ref{fig:fig3} shows the crossover from $\Gamma\ll k_B T$ to $\Gamma\gg k_B T$, illustrating several qualitative features.  The $k_B\ln{3}$ peak in $\Delta S(\mu)$ decreases with $\g$, until no excess entropy is visible at the charge degeneracy point for $\Gamma/k_B T\gtrsim 5$ (Figs.~\ref{fig:fig3}a). This suppression of the entropy associated with charge degeneracy originates from the broadening by $\Gamma$ of the $N=1$ level due to hybridization with the continuous density of states in the reservoir \cite{Sela.2019}.  At the same time, the total entropy change $\Delta S_{0\rightarrow 1}$ remains $\sim k_B\ln{2}$ over the entire range of $\g$ explored in this experiment, reflecting the entropy of the spin-1/2 electron trapped in the QD.

To make quantitative comparison between theory and experiment, we employ NRG simulations \cite{Legeza.2008, Toth.2008}, that yield $N$ as a function of both $T$ and $\eo$, where $-\eo$ is the depth of the dot level below the reservoir potential.   From $N(T,\eo)$, $dN/dT$ and therefore $\Delta S$ may be extracted via Eq.~\ref{eq:eran}. In order to make a direct comparison with the experiment, $\Delta\eo\equiv\eo-\eo(N=1/2)$ is defined like $\Delta V_D$, centred with respect to the charge transition.  
NRG parameters are calibrated to match those in the measurements by aligning the occupation $N(\Delta \epsilon_0$) with the measured $N(\Delta V_D$)\cite{supplement}, from which the appropriate $\Gamma/T$ calculation may be selected and the precise connection between $\Delta \epsilon_0$ with $\Delta V_D$ is ensured. As seen in Fig.~\ref{fig:fig3}b, the data/theory agreement is perfect in terms of dot occupation, within the experimental resolution, giving confidence that measured and calculated $\Delta S$ may be compared directly.

The calculations in Fig.~\ref{fig:fig3}c illustrate NRG predictions for $\Delta S(\eo)$  over the range of \g~accessible in our measurements. Matching the data, the peak in entropy due to charge degeneracy is suppressed as \g~increases above $\g\gtrsim k_B T$, while the net entropy change across the transition remains $k_B \ln 2$.  At the same time, a clear qualitative difference between data and NRG is the shift to the right seen in NRG curves for higher \g~ (Fig.~\ref{fig:fig3}c), but not observed in the measurements (Fig.~\ref{fig:fig3}a).  
This relative shift of NRG with respect to data is not explained by an offset of $\Delta \epsilon_0$ with respect to $\Delta V_D$, as the two are aligned by the occupation data (Fig.~\ref{fig:fig3}b).

At first glance, the shift of NRG curves to the right (to larger chemical potential) with increasing $\Gamma$ may seem strange: why would the emergence of $k_B \ln 2$ entropy in the dot be delayed until after the electron is mostly localized in the dot?  This behaviour is explained by the virtual exchange interactions underlying the Kondo effect, which form a quasi-bound singlet state between the localized spin and a cloud of delocalized spin for temperatures below $T_K$.  This results in a state with no magnetic moment \cite{Kondo.1964} and, in the case of a single-electron QD, zero entropy. Thus, due to the Kondo effect, we expect the entropy to remain zero for all chemical potentials that obey $T<T_K(\mu)$. 
Since $T_K\propto e^{-\pi(\eo-\mu)/\Gamma}$ in the (experimentally relevant) large-$U$ limit, where $U$ represents the QD charging energy, we expect the onset of $k_B$ln 2 entropy to shift to larger values of $\mu$ as $\Gamma$ increases, as seen in the NRG results. (Note that NRG curves are independent of the values of bandwidth, $W$, and of $U$, as long as $U,W\gg T,\Gamma$ because, as with the experimental plots, all curves are shifted such that $x=0$ corresponds to half filling.)

For the present device geometry, $\Delta I_{CS}$ collapses with increasing $\Gamma$, limiting the strength of the coupling accessible in the measurement to $\g/k_B T\lesssim 25$ and preventing us from reaching the deep Kondo regime far from the charge transition.  $T\ll T_K$ may still be achieved close to the charge transition, however, in the so-called mixed-valence regime where $|\eo-\mu|\lesssim\g$ and the electron is not yet strongly localized in the dot.
It remains a puzzle why the strong suppression of entropy right at the charge transition, seen in NRG calculations for $\Gamma/k_B T\geq 5$, is not observed in the data.

It has been suggested \cite{silva2003peculiarities}  that the charge measurement itself can lead to dephasing of the Kondo singlet. Ref.~\onlinecite{avinun2004controlled} reported a suppression of the Kondo peak in the conductance due to charge sensing.  Those results implied  stronger dephasing than had been predicted theoretically, and led to new theoretical work proposing alternative dephasing mechanisms\cite{PhysRevLett.95.206808}. In order to test for charge-sensor dephasing in our measurement, the experiment was repeated at $V_{CS}$ down to 50 $\mu$V, but no dependence on $V_{CS}$ was seen in the data \cite{supplement}.  The discrepancy between NRG and measurements therefore points to the relevance of other degrees of freedom, such as other single-electron levels, that have not been included in current theories.  In the future, experiments that allow simultaneous transport and entropy characterization of the Kondo state may help to resolve this puzzle.  We note that the entropy measurement presented here is the first that could be sensitive to the dephasing of the Kondo state itself, rather than dephasing of transport through the Kondo resonance\cite{kang2007entanglement}.

ACKNOWLEDGEMENTS: This project has received funding from European Research Council (ERC) under the European Union’s Horizon 2020 research and innovation program under grant agreement No 951541. Y. Meir acknowledges discussions with A. Georges and support by the Israel Science Foundation (grant 3523/2020).  Experiments at UBC were undertaken with support from the Stewart Blusson Quantum Matter Institute, the Natural Sciences and Engineering Research Council of Canada, the Canada Foundation for Innovation, the Canadian Institute for Advanced Research, and the Canada First Research Excellence Fund, Quantum Materials and Future Technologies Program.  S.F., G.C.G. and M.M. were supported by the US DOE Office of Basic Energy Sciences, Division of Materials Sciences and Engineering award DE-SC0006671 and QIS award DE-SC0020138. AKM acknowledges funding from the Irish Research Council Laureate Awards 2017/2018 through grant IRCLA/2017/169.

\bibliography{draft}

\widetext
\clearpage
\newpage
\setcounter{equation}{0}
\setcounter{figure}{0}
\setcounter{table}{0}
\setcounter{page}{1}
\renewcommand{\theequation}{S\arabic{equation}}
\renewcommand{\thefigure}{S\arabic{figure}}
\makeatletter
\renewcommand{\present@bibnote}[2]{}
\makeatother

\input{supplement}
\end{document}

%% file: supplement.tex
\newcommand{\dt}{\ensuremath{\Delta T}}
\newcommand{\dn}{\ensuremath{\Delta N}}

\section{Device Fabrication}
The device was fabricated in a GaAs/AlGaAs heterostructure that hosts a 2D Electron Gas (2DEG) 57nm below the surface and that had a 300 mK carrier density of $2.42\times10^{11}\text{cm}^{-2}$ with mobility $2.56\times10^6\text{cm}^2/(\text{Vs})$. A UV laser writer was used to define the mesas, followed by electron beam lithography to define NiAuGe ohmic contacts. Additionally, 10 nm of HfO\textsubscript{2} was deposited by atomic layer deposition to improve gating stability. The electrostatic gates were fabricated with two stages of electron beam lithography followed by electron beam evaporation: a fine step for the inner parts, and a coarse step for the outer parts of the gates. In the fine step, 2/12 nm of Pd/Au were deposited. In the coarse step, 10/150 nm of Ti/Au were deposited. 

\section{Measurement Electronics}
A custom built combined DAC/ADC unit was used to apply potentials to the gates and heating QPCs as well as to record the voltage output of a Current to Voltage Basel SP983c amplifier (https://www.baspi.ch/low-noise-high-stab-itov-conv). The DAC/ADC unit is built from an Arduino Due and two Analog Devices evaluations boards: the AD5764 DAC and AD7734 ADC. The Arduino is the interface between the measurement PC and the DAC/ADC boards. The whole design is based on the information provided at [http://opendacs.com/dac-adc-homepage/] with some substantial modifications, particularly to the Arduino code (https://github.com/folk-lab/FastDAC). The most significant modification is to provide functionality to apply a synchronized square wave bias for heating, whilst measuring continuously.

\section{Charge sensor mapping}
\begin{figure}[h]
    \centering
    \includegraphics{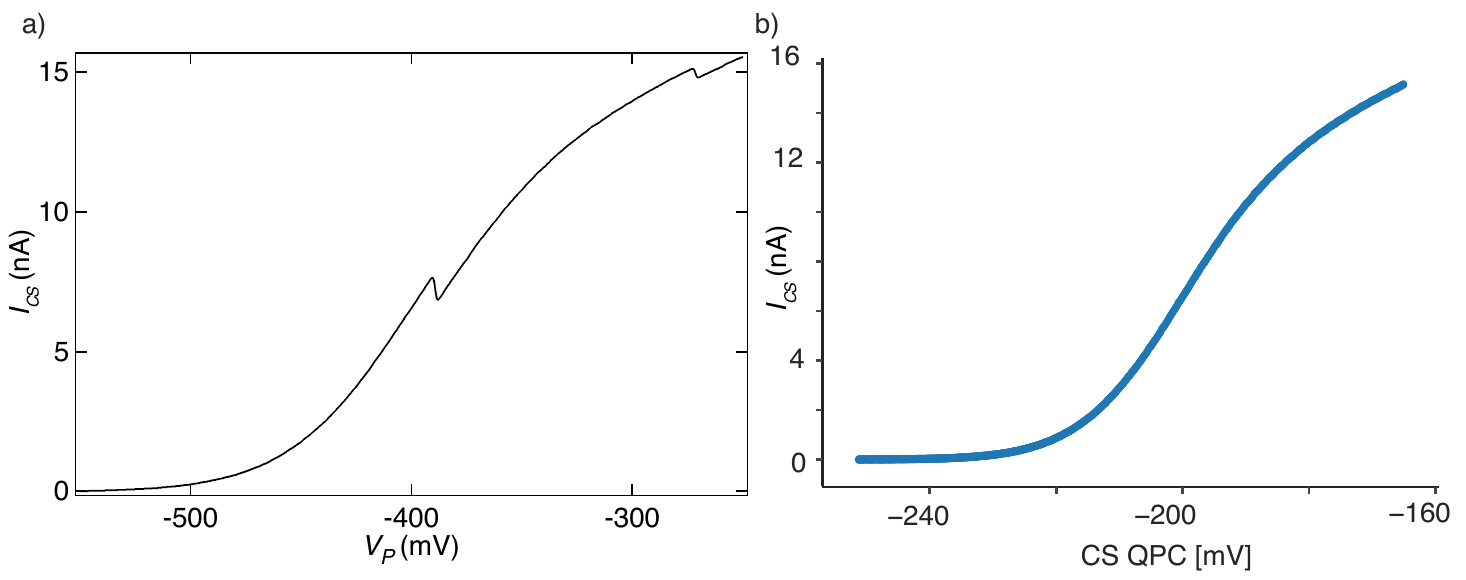}
    \caption{
    a) A very wide sweep over a weakly coupled $0 \to 1$ transition using the plunger gate showing the response of the charge sensor, and the usual alignment of the $0 \to 1$ transition at the steepest, most linear part of the charge sensor response.  b) Data of charge sensor current, \is, vs the charge sensor QPC tuning gate (see Fig.~\ref{fig:fig1}a) used for mapping \is to a  quantity in units of mV that is linear in electrostatic potential.
    }
    \label{fig:sup_cstrace}
\end{figure}

The charge sensor is tuned to the most linear regime before each measurement (Fig.~\ref{fig:sup_cstrace}a).  In the limit of very strong coupling, however, the transition becomes so broad that the non-linearity of the charge sensor may begin to play a role.  In our experiment, $\Delta\is$ is converted to $\Delta N$ assuming a linear relation between the two, but when $\is$ is not linear in the additional electrostatic potential provided by, e.g., cross capacitance with $V_D$, this assumption is no longer valid.

To remove potential inaccuracy in the conversion between $\Delta\is$ and $\Delta N$ due to non-linearity, \is~may be mapped back to an equivalent charge sensor gate voltage using a measurement  $\is(V_{QPC}$ (Fig.~\ref{fig:sup_cstrace}b). Performing the entropy calculation using the equivalent gate voltage, instead of \is, significantly reduces any impact of charge sensor non-linearity in the measurement of $\Delta N$.  In practice, however, no statistically significant difference was observed in entropy calculations using the two approaches, so charge sensor mapping was not used.

\section{Determining $\Delta S$ by direct fitting of \dndt~data}

For systems where the addition of an electron to the charge sensed quantum dot has a simple lineshape, as in the case of a QD weakly coupled to a thermal reservoir, it is possible to extract the entropy change of the system by fitting the \di~data directly as was done in Ref. \onlinecite{Hartman.2018}. This procedure is made possible through the application of the Maxwell relation:
$$
\left ( \frac{\partial \mu}{\partial T} \right )_{p,N} = - \left ( \frac{\partial S}{\partial N} \right )_{p, T}
$$

\noindent For a QD weakly coupled to a thermal reservoir, the charging lineshape takes the form:
$$
N(\vp, \Theta) = \tanh \left( \frac{\vp-V_{mid}(\Theta)}{2\Theta}\right)
$$
where $\Theta = \frac{k_B T}{\alpha e}$ and $V_{mid}(\Theta)$ is the plunger gate voltage at $N = 1/2$.  $V_{mid}$ changes the quantum dot energy level--that is, the energy required to add an electron to the dot--and therefore maps to the chemical potential $\mu$ in equilibrium.

Differentiating $N$ with respect to $T$, one finds a lineshape that depends explicitly on $\Delta S$:
$$
\partial N(\vp, \Theta) \propto -\partial T \left[\frac{\vp - V_{mid}(\Theta)}{2\Theta} - \frac{\Delta S}{2k_B}\right] \times \cosh^{-2}\left(\frac{\vp - V_{mid}(\Theta)}{2\Theta}\right)
$$
 after substituting $$\frac{\partial V_{mid}}{\partial \Theta} = \frac{1}{k_B}\frac{\partial \mu}{\partial T} = \frac{1}{k_B}\Delta S_{N-1 \rightarrow N}$$.
 
Here $\partial N(\vp, \Theta)$ is the difference in occupation of the QD for the reservoir temperature changing from $T \rightarrow T + \Delta T$. Fitting this equation to the \di~ data, $\Delta S$ is obtained as a fit parameter independent of the scaling of the data. 

\section{Fitting NRG to Data}
NRG calculations were carried out using the flexible DM-NRG code\cite{Legeza.2008,Toth.2008} on the standard single impurity Anderson model. In the calculation, we assume infinite interaction $U$, a constant density of states in the reservoir with bandwidth $D=1$ exceeding all other energy scales, and we keep 350 states per iteration with discretization constant $\Lambda=2$. Results are given in arbitrary units of energy. 
Our procedure for fitting the occupation, $N$, of NRG calculations to measured data involves three steps: 1. Linearly interpolating over the 2D array of calculations. 2. Adding terms (amplitude, constant, linear) to account for the behaviour of the charge sensor in detecting the QD occupations. 3. Allowing for an offset and scaling proportional to $\Theta$ ($T$ in units of gate voltage) in the \eo~ axis. We then use Powell's method of minimization \cite{Powell.1964} to find the best fitting parameters allowing all to vary with the exception of \g~ and $\Theta$, for which only one is allowed to vary. In the weakly coupled regime, it is reasonable to approximate $\g \sim 0$, and with that constraint, we are able to determine $\Theta(\vp)$. We find a linear relationship between $\Theta(\vp)$ and \vc~ which implies a linearly changing lever arm, $\alpha$, as the system temperature, $T$, is fixed. Note that the lever arm, $\alpha$, connects $\Theta$ in units of gate voltage to temperature, $T$, in kelvin (Eq.\ref{eq:alpha}),
\begin{equation}\label{eq:alpha}
   \alpha \Theta = k_B T 
\end{equation}
\noindent and is a measure of the strength of effect the plunger gate, \vp, has on the QD. The linear change implies that as \vc~ is varied, the strength of effect of \vp~ also varies. We attribute this to a change of shape of the QD where it moves it further from \vp~ for more positive \vc. For measurements into the strongly coupled regime where $\g \gg 0$, we force the $\Theta$ parameter to follow the linear relationship found in the weakly coupled regime, allowing \g~ to be a varying parameter. The fit parameters found by comparing $N$ of NRG data to \is~ of measured data can then be used to directly compare between the NRG \dndt~ calculations and \di~ measurements.

\section{Scaling from \di~to \dndt}
The complete procedure for scaling from \di~to \dndt~is comprised of two parts: Conversion of \di~to \dn, and calculation of the corresponding \dt, expressed in equivalent mV on $V_D$. 

The procedure for scaling the \di~measurements to \dndt~involves scaling $\di \to \dn$, then dividing by \dt~as described in the main text. 
The $\di \to \dn$~conversion is a straightforward division by $I_e$, the net change of current through the charge sensor for the addition of 1 full electron to the QD. Extracting $I_e$ from the data is achieved by scaling NRG calculations of occupation across the transition to the heated portion of $I_{CS}$ data, with additional offset and linear terms added to the NRG to account for cross capacitance between $V_D$ and the charge sensor. 

\dt~is easily extracted in units of  equivalent gate voltage (\vp) for weakly coupled $\vc$ by fitting cold and hot occupation data to NRG. For strongly coupled transitions, however, \dt~does not result in a broadening of the transition lineshape, so it must be determined in another way. The real temperature change of the reservoir does not depend on \vc, of course, but the lever arm $\alpha$ does depend on \vc.  We calculate \dt(\vc) in equivalent mV on $V_D$ by
\begin{enumerate}
    \item fitting hot and cold transitions for a range of weakly coupled \vc, to determine both $\Delta T(\vc)$ in equivalent $V_D$ and $\alpha(\vc)$ through this range.
    \item $\alpha(\vc)$ is observed to be linear in \vc, and extrapolated to strongly-coupled  \vc~(dashed line in Fig.~\ref{fig:fig1}c, main text).
    \item $\Delta T$ in equivalent $V_D$ is calculated for strongly coupled transitions using $\alpha(\vc)$ determined above.
\end{enumerate}

\newpage

\section{Averaging Procedure}
The data shown in the main text is the result of averaging measurements over many sweeps over the transition. In the strongly coupled regime,  as the $\Delta I_{CS}$ signal becomes weaker, averaging data becomes particularly important.  It is often necessary to measure for 10's of minutes or even hours in order to obtain a reasonable signal to noise ratio, but the presence of charge instability makes single slow measurements over the transition unreliable. By repeatedly sweeping over the transition quickly, then aligning each sweep based on a fit to the $I_{CS}$ data before averaging, we can improve the signal to noise ratio of the corresponding $\Delta I_{CS}$ whilst mitigating the effect of charge instability.  This procedure of post-aligning individual charge transition scans is followed for measurements in the range $V_T < -230$ mV.

For more strongly coupled measurements ($V_T\ge -230$ mV), determining the center of each individual scan through fitting is not reliable; as a result, data is averaged without centring first. Charge instability (determined in the weakly coupled regime) is on the order of 6 $\mu$eV without significant long-term drift seen in the data. For the most strongly coupled measurements, where the width of the transition is on the order of 400 $\mu$eV, the lack of centring therefore is expected to have a negligible effect. Occasional larger jumps in transition position (~0.5 mV) do occur on a timescale of hours; care is taken never to average data across such jumps. 

\begin{figure}[h]
    \centering
    \includegraphics{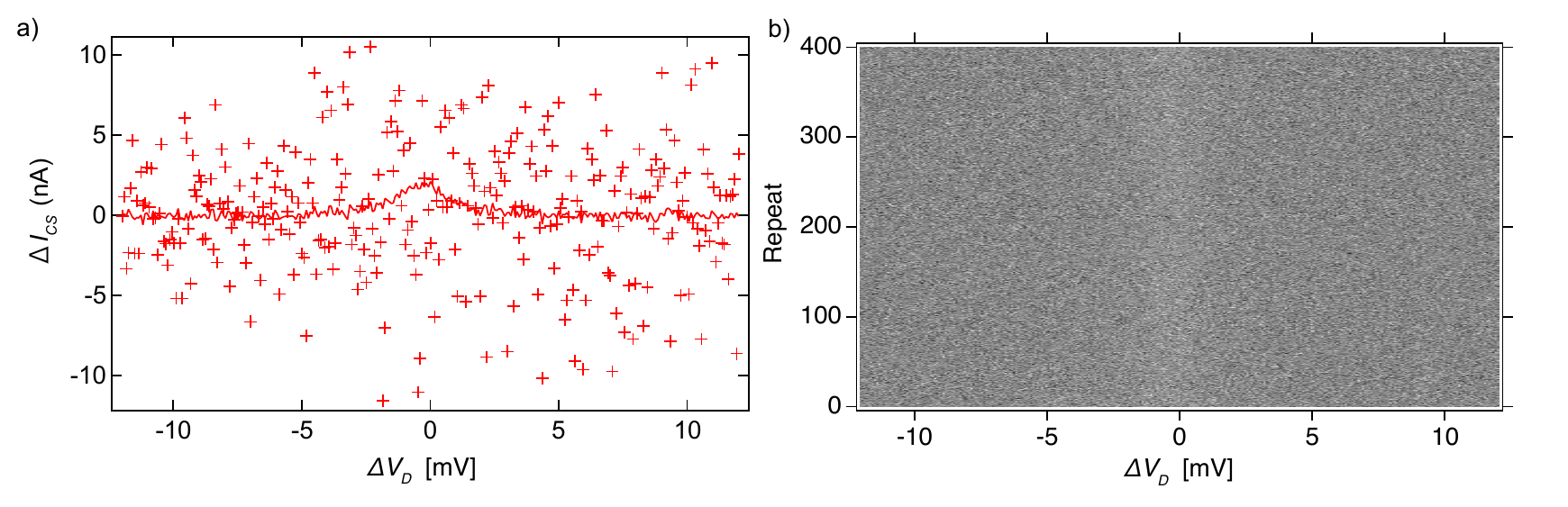}
    \caption{
    a) Markers illustrate a single measurement across the charge transition, which takes 30 seconds to complete.  No peak in $\Delta\is$ can be seen in this raw data.  After averaging 400 of such scans together (solid line), however, a small peak in $\Delta\is$ is seen at $\Delta V_D=0$. b) Raw $\Delta\is$ data (greyscale) for 400 scans as in panel a). Averaged together, they yield the solid line in panel a).
    }
    \label{fig:single_gamma}
\end{figure}

\newpage

\section{Lack of dependence on charge sensor bias}

\begin{figure}[h]
    \centering
    \includegraphics{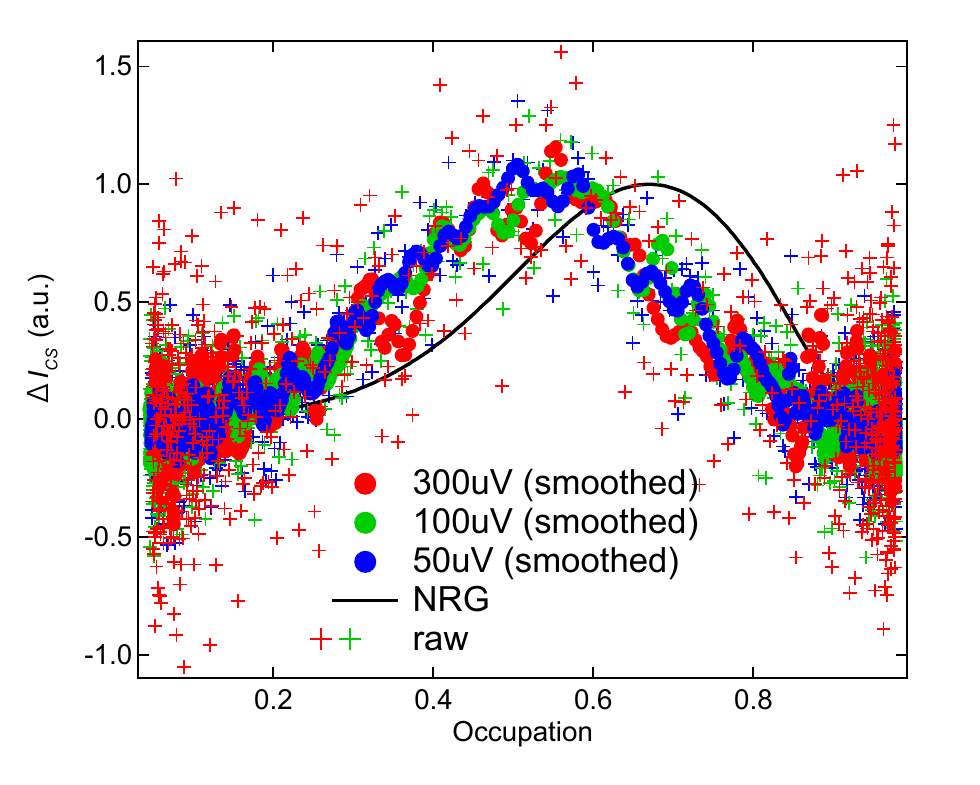}
    \caption{
    The lineshape of $\Delta\is$, here plotted vs occupation instead of $V_D$, shows no dependence on $V_{CS}$ within experimental noise, though of course the  magnitude  of $I_{CS}$ and $\Delta I_{CS}$ scales linearly with $V_{CS}$.  The case of $\Gamma/k_B T = 24$ is shown here. In particular, $\Delta\is$ remains peaked at $N\sim 0.5$, in contrast to the NRG calculation (solid line) in which the shifted peak reflects the screening of spin entropy in the mixed valence regime due to the formation of the Kondo singlet.    
    }
    \label{fig:my_label}
\end{figure}